\documentclass[a4paper,fleqn, 10pt]{article}

\usepackage{geometry}

\usepackage[british]{babel}
\usepackage{amsthm, amssymb, mathtools, amsmath}
\usepackage[dvipsnames,svgnames,x11names]{xcolor}
\usepackage{amsbsy}
\usepackage{enumerate}
\usepackage{mdwlist}
\usepackage{bigints}

\usepackage{enumitem}

\usepackage[numbers]{natbib}

\usepackage{todonotes}
\DeclareMathOperator{\Exp}{\mathbb{E}}

\usepackage{graphicx}
\usepackage{subfigure}
\usepackage{float}

\usepackage[normalem]{ulem} 

\usepackage[labelfont=bf,font=small,up]{caption}

\usepackage{lscape} 

\usepackage{url, verbatim}

\usepackage{etoolbox}
\makeatletter
\patchcmd{\maketitle}{\@fnsymbol}{\@arabic}{}{}  
\makeatother

\title{Who is the infector? Epidemic models with symptomatic and asymptomatic cases}
\date{\today}
\author{Ka Yin Leung$^1$ \and Pieter Trapman$^1$ \and Tom Britton$^1$}

\begin{document}
\maketitle

\begin{abstract}
What role do asymptomatically infected individuals play in the transmission dynamics? There are many diseases, such as norovirus and influenza, where some infected hosts show symptoms of the disease while others are asymptomatically infected, i.e.\ do not show any symptoms. The current paper considers a class of epidemic models following an SEIR (Susceptible $\to$ Exposed $\to$ Infectious $\to$ Recovered) structure that allows for both symptomatic and asymptomatic cases. The following question is addressed: what fraction $\rho$ of those individuals getting infected are infected by symptomatic (asymptomatic) cases? This is a more complicated question than the related question for the beginning of the epidemic: what fraction of the expected number of secondary cases of a typical newly infected individual, i.e.\ what fraction of the basic reproduction number $R_0$, is caused by symptomatic  individuals? The latter fraction only depends on the type-specific reproduction numbers, while the former fraction $\rho$ also depends on timing and hence on the probabilistic distributions of latent and infectious periods of the two types (not only their means). Bounds on $\rho$ are derived for the situation where these distributions (and even their means) are unknown. Special attention is given to the class of Markov models and the class of continuous-time Reed-Frost models as two classes of distribution functions. We show how these two classes of models can exhibit very different behaviour. 
\end{abstract}


\footnotetext[1]{Department of Mathematics, Stockholm University, 106 91 Stockholm, Sweden.\\ Email: \{tom.britton, kayin.leung, ptrapman\}@math.su.se}

\section{Introduction}
For many known human infectious diseases we have, to some extent, an idea about their clinical features, e.g.\ the typical incubation period and common symptoms associated to them. Despite this, there can be large heterogeneity between infected hosts in a population. Some infected individuals may never show symptoms of disease (i.e.\ are asymptomatic) while others do (i.e.\ are symptomatic). Suppose that we can categorize infected individuals as either asymptomatic or symptomatic, and that some significant fraction of the infected host population is asymptomatic. Then, even if asymptomatics make up a large part of the infected host population, can we say how important their role is in the transmission dynamics? 

In order to answer such a question, one needs to define the meaning of `importance'. A natural approach is to consider the beginning of the epidemic, in particular the basic reproduction number $R_0$. This approach was taken by~\cite{Fraser2004}, where they considered the (weighted) contribution $\theta$ of infectious individuals without symptoms (either asymptomatically infected or prior to symptom onset) to the basic reproduction number. The idea is that if $\theta$ is small, i.e.\ most of the secondary cases in the beginning of an epidemic are caused by symptomatic individuals, then control measures such as isolation of such (in principle detectable) individuals would greatly reduce spreading whereas this approach may not be successful if $\theta$ is large.

In case of an emerging infectious disease such as the recent 2013-2015 Ebola outbreak, one will mostly be interested in (controlling) the beginning of the outbreak. In such cases, we want to understand epidemiological quantities such as $R_0$ and $\theta$. However, in case of infectious diseases such as seasonal influenza in the general population, it is quite usual that major outbreaks occur. In particular, the infectious disease is present in the population beyond its initial phase of the outbreak. In the current paper we are interested in the fraction $\rho$ of the final epidemic size that were infected by symptomatic cases. Understanding the contribution of symptomatic cases to the final size leads to better insights into possible transmission routes, but also how things would change when e.g.\ removing a fraction of symptomatic individuals. 

We consider a simple class of epidemic models that we call the symptom-response SEIR epidemic models, defined in Section \ref{sec:model}. This class of models has an SEIR (Susceptible $\to$ Exposed $\to$ Infectious $\to$ Recovered) structure. The time that an individual spends in either the exposed or infectious period can take on some general distributions. In particular, these distributions need not \emph{necessarily} be exponentially distributed or be the same for symptomatic and asymptomatic cases. We can use the standard toolbox for analyzing mathematical models for infectious disease dynamics to derive e.g.\ the basic reproduction number and the final size. In fact, these and other characteristics have been studied before, e.g.\ in~\cite{BC95}, which we will make use of. 

Characterizing the fraction $\rho$ turns out to be much more complicated than the related fraction $\theta$ that only deals with the beginning of the outbreak on a generation basis. Whereas $\theta$ only depends on the type-specific reproduction numbers (i.e.\ the expected number of secondary cases generated by a newly infected individual that is either symptomatic or asymptomatic), the fraction $\rho$ also depends on the mean latent and infectious periods of the two types as well as their probabilistic distributions. The general characterization of $\rho$, given in Section~\ref{sec:cause}, does not provide much insight even though it may be computed numerically when all distributions and parameters are given. Because of this, and the fact that distributions of latent and infectious periods are rarely fully known, it is of interest to give upper and lower bounds for $\rho$. In a `twin' paper~\cite{guilty} we use probabilistic tools to derive bounds for $\rho$ for the setting where these distributions (including their means) are unknown (but note that~\cite{guilty} is more general than determining the upper- and lower bound for the symptom-response SEIR model). In Section~\ref{sec:Bounds} of the current paper, we present these bounds and explain the heuristics in deriving them. Moreover, we are interested in two special classes of the model, namely the class of Markov models with exponentially distributed latent and infectious periods and the class of continuous-time Reed-Frost models with constant latent and infectious periods. We show that the choice of distribution functions can have significant qualitative differences on the time-evolution of the epidemic by comparing the class of Markov models with the Reed-Frost models having identical means. Furthermore, in Section 4 we consider different scenarios in order to gain insights into $\rho $ through numerical investigation and to illustrate numerically the qualitative differences between Markov models and Reed-Frost models.

\section{Model definition}\label{sec:model}

\subsection{The stochastic symptom-response SEIR epidemic model}\label{sec:model_stoch}
The \emph{stochastic symptom-response SEIR epidemic model} is defined as follows (in our twin paper~\cite{guilty} we study a more general form of the current model). We consider a closed population of homogeneously mixing individuals, all being equally susceptible. We let $n$ denote the number of individuals in the population. The specific feature of the model is that infected individuals can either become \emph{symptomatic} or \emph{asymptomatic}. In order not to confuse \emph{S}ymptomatic with \emph{S}usceptible, we use subscript $d$ to denote symptomatic individuals, i.e.\ the individuals that show signs of the \emph{d}isease. Individuals are initially susceptible (S). If an individual gets infected it becomes symptomatic with probability $p_d$ and asymptomatic with the remaining probability $p_a=1-p_d$. These events are independent between individuals and also independent of whom the individual was infected by (see~\cite{Ball2007} for a situation where the response does depend on the infector).

We let $S(t)$ denote the number of individuals in the population that are \emph{S}usceptible at time $t$. If the individual becomes symptomatic, it first has a random latency period $L_d$ (and we let $E(t)$ with a subscript $a$ or $d$ denote the number of asymptomatic and symptomatic individuals in the population at time $t$ that are infected and \emph{E}xposed but not yet infectious) followed by a random infectious period $\iota_d$ (where we let $I(t)$ with subscripts $a$ and $d$ denote the number of \emph{I}nfectious individuals that are asymptomatic and symptomatic). Symptomatic individuals have infectious contacts at rate $\lambda_d$ during the infectious period, each time with a uniformly chosen individual in the population. Infectious contacts with susceptible individuals result in the latter getting infected. Other infectious contacts have no effect on the epidemic (hence the term \emph{infectious} contacts). The situation for asymptomatic cases is analogous with latency period $L_a$, infectious period $\iota_a$, and infectious contact rate $\lambda_a$. When the infectious period of an individual terminates, the individual recovers and becomes immune (where $R(t)$ with subscript $a$ and $d$ denotes the number \emph{R}ecovered individuals that were asymptomatic or symptomatic) for the rest of the epidemic. Note that $S(t)+E_d(t)+E_a(t)+I_d(t)+I_a(t)+R_d(t)+R_a(t)=n$. The epidemic starts with a small number of infectious individuals and all other individuals susceptible. The end of the epidemic is at time $T$, where $T$ is the first time that there are no more latent or infectious individuals around. At time $T$ there are only susceptible and recovered individuals. Let $Z_d=R_d(T)$ and $Z_a=R_a(T)$ be the final number (previously) symptomatic and asymptomatic cases, respectively (so $S(T)=n-(Z_d+Z_a)$). The overall number of individuals infected is denoted by $Z=Z_d+Z_a$. We are mainly interested in large population sizes and hence study the situation where $n\to\infty$. Moreover, we let $\bar S(t)=S(t)/n$ and we define other population quantities decorated with a bar over it in an analogous fashion. The limit of $Z_d/Z$ as $n\to\infty$ conditioning on $Z\to\infty$ is simply the probability $p_d$ that a newly infected individual becomes a symptomatic case, and $Z_d/n\to p_dz$, where $z$ is the large population limit fraction of the population getting infected in case of a major outbreak (and is characterised by the final size equation~\eqref{eq:finalsize} below).

Note that we have not specified anything about symptoms and when they appear. In fact, we could equally well model a situation with mildly infected and severely infected (or two types 1 and 2 for that matter; see also~\cite{guilty}). Having symptomatic cases in mind, the contact rate $\lambda_d$ should reflect both the more likely situation that symptomatic cases have a higher viral load as compared to asymptomatic cases (thus leading to higher transmission probability upon contact), and the fact that symptomatic cases may reduce their number of social contacts and thus meet fewer individuals. Since the infectious contact rate reflect both components, it need not necessarily be that $\lambda_d>\lambda_a$. 

Note that if there are no latency periods ($L_d\equiv L_a\equiv 0$) the model reduces to an SIR model. Furthermore, we give special attention to two distributions for latent and infectious period when studying S(E)IR models. The first class of distributions is where both latency and infectious periods follow exponential distributions, giving rise to Markov models. In the deterministic setting this corresponds to the model being described by ordinary differential equations (ODE). The second class of distributions is where these periods are non-random, i.e.\ fixed and the same for all symptomatic (asymptomatic) individuals. We refer to this class of models as continuous-time Reed-Frost epidemic models, since the events to infect different individuals are then independent. In the deterministic setting such models can be described with integral equations.

\subsubsection{Basic reproduction numbers and the final size}\label{sec:epi}
If we do not distinguish symptomatic and asymptomatic cases, the model defined in Section~\ref{sec:model_stoch} falls under the general framework of epidemic models defined in~ \cite{Ball1986}, and treated specifically in~\cite{BC95}. The fundamental quantity is the distribution of $C$, the (random) number of infectious contacts that an infected individual has during its infectious period. To this end, let the random variables $C_d$ and $C_a$ denote the number of infectious contacts for symptomatic and asymptomatic individuals, respectively. 

The final size distribution is completely determined by the distribution of $C$ (which in turn is specified by $C_d$, $C_a$ and $p_d$). For the symptom-response SEIR model, $C$ follows a mixed Poisson distribution: $\text{MixPo}\big(\lambda_d1_d\iota_d+\lambda_a(1-1_d) \iota_a\big)$ (all random variables being independent and not necessarily exponentially distributed), where $1_d$ is 1 with probability $p_d$ and 0 otherwise. Note that the latency periods $L_a$ and $L_d$ have no effect on the \emph{number} of infections -- they only affect the \emph{timing} of the outbreak.   

The basic reproduction number denotes the expected number of infectious contacts and equals 
\begin{equation}\label{eq:R0}
R_0=\Exp(C)=p_d\Exp(C_d) + (1-p_d)\Exp(C_a) = p_d\lambda_d\Exp(\iota_d) + (1-p_d)\lambda_a\Exp(\iota_a)= p_dR_{0,d}+(1-p_d)R_{0,a},
\end{equation}
where we call $R_{0,d}=\lambda_d\Exp(\iota_d)$ and $R_{0,a}=\lambda_a\Exp(\iota_a)$ the type-specific reproduction numbers for newly infected symptomatic and asymptomatic individuals, respectively. We denote the fraction of $R_0$ that is caused by symptomatic individuals by $\theta_d$, so
\begin{equation}\label{eq:theta}
\theta_d=\frac{p_dR_{0,d}}{R_0}.
\end{equation}
(Note that~\cite{Fraser2004} consider the quantity $\theta=1-\theta_d$ for a related but different model.) 

Next, we make use of known results on the final size $Z$ from~\cite{BC95}. First of all, if $R_0>1$, then the final fraction infected $\bar Z=Z/n$ converges to a two-point distribution with probability mass at 0 (small outbreaks) and at another point $z$, where $z$ is the unique positive solution to the equation 
\begin{equation}\label{eq:finalsize}
1-z=e^{-R_0z}
\end{equation} 
(major outbreaks). So, if $R_0\le 1$ only minor outbreaks are possible. The probability for a major outbreak depends more delicately on the distribution of $C$, and not only on its mean (e.g.~\cite{BC95}). Second of all, in case of a minor outbreak, $Z$ has a limiting distribution described by the finite part of the distribution of the total progeny of a branching process with offspring distribution $C$. Moreover, there is a central limit theorem for $\bar Z$ in case of a major outbreak. 

We use these results on the final size in the present paper. However, our focus is different: we are interested in the number of cases that got infected by symptomatic and asymptomatic individuals, respectively. Here, we let $Y_d$ and $Y_a$ denote the number of individuals that got infected by a symptomatic and asymptomatic case, respectively (so $Z=Y_a+Y_d$). We are interested in the law of large number limit $\rho_d$ that denotes the fraction of infected individuals that get infected by symptomatic cases, given a major outbreak. We hence seek the limit $\rho_d$ of $Y_d/Z$ as $n\to\infty$ conditioning on $Z\to\infty$ (note the symmetry in asymptomatic and symptomatic cases, $\rho_a+\rho_d=1$, so we can just as well switch the roles of $a$ and $d$). We turn to this in Section~\ref{sec:cause}. 

\subsection{Deterministic symptom-response epidemic models}\label{sec:model_det}
As with nearly all epidemic models, the current model may be approximated by a deterministic epidemic model when the population size is large enough and when the number of individuals in the different compartments are large enough. More specifically, when $n$ is large, $(\bar S(t), \bar E_a(t), \bar E_s(t), \bar I_a(t), \bar I_d(t), \bar R_a(t), \bar R_d(t))$ may be approximated by a deterministic model on the part of the time axis where all fractions are not too close to 0. The latter means that the approximation does not work in the beginning nor at the end of an outbreak (when the \emph{number} of latent and infectious individuals are moderate to small), since these settings imply that the corresponding fractions are close to 0. If the epidemic is initiated by a small positive \emph{fraction} of infected people (either latent or infectious and symptomatic or asymptomatic), the starting phase is removed. In that case the approximation applies on any fixed time interval $[0,t]$. 

The statement that the stochastic model may be approximated by a deterministic counterpart really should mean that there is a law of large number theorem that proves that, as $n\to\infty$, the stochastic process for the fractions converge in probability, uniformly on bounded intervals, to the deterministic process. It is not the purpose of the present paper to prove such results for the current model. Instead we assume this to hold and therefore work interchangeably with the deterministic and stochastic setting. The law of large numbers should be proven using population process theory (e.g.\ Ethier and Kurtz, 2006) in ways similar to what has been done for related models, cf.~\cite{Diekmann2013}, Section 3.4.

In the deterministic formulation of the model, we let the population size $n\to\infty$ and consider expected fractions of the population in different disease states. The corresponding population quantities are denoted by a small letter, i.e.\ $s(t)$ is the expected fraction of the population that is susceptible at $t$, and similarly for other population quantities.  

We do the bookkeeping using the age $\tau$ since infection, taking into account asymptomatic and symptomatic cases. Let 
\begin{equation}\label{eq:inf_prob}
\begin{aligned}
\pi_a(\tau)&\coloneqq P(L_a\leq \tau\leq L_a+\iota_a)
\end{aligned}
\end{equation}
i.e.\ $\pi_a(\tau)$ denotes the probability that an asymptomatic individual is infectious at age since infection $\tau$. Furthermore, let $F_a(t)$ denote the force of infection from asymptomatic individuals at time $t$ (and similarly we have $\pi_d(\tau)$ and $F_d(g)$ for symptomatics). So, if $i_a(t)$ and $i_d(t)$ denote the fraction of individuals that are asymptomatic infectious and symptomatic infectious respectively, then $F_a(t)=\lambda_ai_a(t)$ and $F_d(t)=\lambda_d i_d(t)$. Furthermore,
\begin{equation}\label{eq:s}
\dot s=-(F_a+F_d)s.
\end{equation}
From~\eqref{eq:s} it follows that the expected incidence at time $t-\tau$ is $(F_a(t-\tau)+F_d(t-\tau))s(t-\tau)$. Of this quantity, a fraction $p_a=1-p_d$ is asymptomatic. An asymptomatic individual that was infected time $t-\tau$ ago has expected infectivity $\lambda_a\pi_a(\tau)$ at time $t$. Then, by integrating over all possible ages since infection $\tau\geq0$, we find an expression for $F_a(t)$ (and similarly for $F_d(t)$):
\begin{equation}\label{eq:FOI}
\begin{aligned}
F_a(t)&=\int_0^\infty \lambda_a\pi_a(\tau) (1-p_d)\big(F_a(t-\tau)+F_d(t-\tau)\big)s(t-\tau)d\tau\\
F_d(t)&=\int_0^\infty \lambda_d\pi_d(\tau) p_d\big(F_a(t-\tau)+F_d(t-\tau)\big)s(t-\tau)d\tau.
\end{aligned}
\end{equation}
Together,~\eqref{eq:s} and~\eqref{eq:FOI} form a closed system. From this system, one can obtain e.g.\ the mean fraction $e_A(t)$ of the population that is asymptomatically infected and still in the latent period. To this end, we simply note that an asymptomatic individual that was infected time $t-\tau$ ago is in its latent period with probability $P(L_a>\tau)$. Therefore, the fraction of exposed asymptomatic-to-be individuals is
\begin{equation*}
e_a(t)= \int_0^\infty P(L_a>\tau) (1-p_d) \big(F_a(t-\tau)+F_d(t-\tau)\big)s(t-\tau)d\tau,
\end{equation*}
and similarly one can recover the formulas for the other population quantities. 

\subsubsection{Markov models}\label{sec:Markovmodel}
Often in the modelling community, ordinary differential equations (ODE) are used to describe deterministic compartmental disease models. Implicitly this assumes that an individual spends an exponentially distributed amount of time in each disease compartment, giving rise to the special case of Markov models. Assume that $L_k\sim\exp(\alpha_k)$, $\iota_k\sim\exp(\gamma_k)$, $k=a,d$. Then, by working out the probability $\pi_k$, $k=a,d$, to be infectious at age-since infection $\tau$:
\begin{equation*}
\begin{aligned}
\pi_k(\tau)&=\frac{\alpha_k}{\gamma_k-\alpha_k}\left(e^{-\alpha_k\tau}-e^{-\gamma_k\tau}\right). 
\end{aligned}
\end{equation*}
In ODE formulation, the dynamics of the model are described by
\begin{equation}\label{eq:ODE_exp}
\begin{aligned}
s'&= -(\lambda_a i_a+\lambda_d i_d) s\\
e_k'&= p_k(\lambda_a i_a+\lambda_d i_d) s-\alpha_k e_k\\
i_k'&= \alpha_k e_k-\gamma_k i_k\\
r_k'&= \gamma_k i_k,
\end{aligned}
\end{equation}
$k=a,d$, with $p_a=1-p_d$. Note that consistency requires $1=s(t)+e_a(t)+i_a(t)+r_a(t)+e_d(t)+i_d(t)+r_d(t)$. As mentioned earlier, the forces of infection satisfy $F_a(t)=\lambda_a i_a(t)$ and $F_d(t)=\lambda_d i_d(t)$. Finally, note that the type-specific reproduction numbers $R_{0,a}$ and $R_{0,d}$ reduce to $R_{0,a}=\lambda_a/\gamma_a$ and $R_{0,d}=\lambda_d/\gamma_d$ (exactly as one would expect from the interpretation). 

\subsubsection{Continuous time Reed-Frost models}\label{sec:RFmodel}
Another special class of models is when each individual spends a deterministic amount of time in each disease compartment. We refer to this class of models as continuous-time Reed-Frost models. This class of models plays an essential role in both this paper and the twin paper~\cite{guilty}. Assume that $L_k\equiv \ell_k$, and $\iota_k\equiv x_k$, $k=a,d$, with $\ell_k, x_k$ nonnegative constants (so all symptomatics have equal lengths of latent and infectious periods, as do all asymptomatics). Then 
\begin{equation*}
\begin{aligned}
\pi_k(\tau)&=\boldsymbol 1_{(\ell_k, \ell_k+x_k)}(\tau), 
\end{aligned}
\end{equation*}
$k=a,d$. This yields the following renewal equations for the forces of infection $F_k$:
\begin{equation}\label{eq:FOI_det}
\begin{aligned}
F_k(t)&=\int_{t-(\ell_k+x_k)}^{t-\ell_k}p_k\lambda_k(F_a(\tau)+F_d(\tau))s(\tau)d\tau,
\end{aligned}
\end{equation}
$k=a,d$ and $p_a=1-p_d$. By differentiating with respect to $t$, we can reformulate~\eqref{eq:FOI_det} as delay differential equations (DDE), and we end up with a system of three DDE for $s$, $F_a$, and $F_d$:
\begin{equation}\label{eq:DDE_RF}
\begin{aligned}
s(t)'&=-(F_a(t)+F_d(t))s(t)\\
F_k(t)'&=p_k\lambda_k\Big\{\big(F_a(t-\ell_k)+F_d(t-\ell_k)\big)s(t-\ell_k)\\
&\phantom{=\ }-\big[F_a\big(t-(\ell_k+x_k)\big)+F_d\big(t-(\ell_k+x_k)\big)\big]s\big(t-(\ell_k+x_k)\big)\Big\},
\end{aligned}
\end{equation}
$k=a,d$ and $p_a=1-p_d$. For given values of $\ell_a, \ell_d, x_a$, and $x_d$ it is straightforward to numerically simulate $s$, $F_a$, and $F_d$. 

Finally, note that type-specific reproduction numbers $R_{0,a}$ and $R_{0,d}$ reduce to $R_{0,a}=\lambda_a x_a$ and $R_{0,d}=\lambda_d x_d$.

\subsection{Initial conditions}
So far, we have not specified initial conditions, i.e.\ how infection is introduced into the susceptible population at the beginning of the epidemic. In general, in a stochastic formulation, one of two options is chosen. There is at time $t=0$, either a positive fraction of infectious individuals, or a fixed number of infectious individuals, as the population size $n\to\infty$. In other words, if we let $\mu_n=n^{-1}m_n$, then, in the first case, $\mu_n\to\mu$ as $n\to\infty$, with $\mu>0$ constant. In the second case, $m_n=m$ is fixed (and therefore $\mu_n\to0$ as $n\to\infty$). In general, the initial conditions can matter for the possible courses that the epidemic can take, see e.g.~\cite{Ball2005}. Implicitly, for the basic reproduction number and the final size of the epidemic, we assume the second setting of a fixed number $m_n=m$ (but approximately the same results apply if $\mu>0$ but very small). 

As we discuss in Section~\ref{sec:model_det}, the stochastic model of Section~\ref{sec:model_stoch} can be approximated by a deterministic epidemic model when the population size $n$ is large enough. In the deterministic formulation we implicitly assume an infinite population size and work with mean fractions of the population. In particular, if we define the initial condition at $t=0$, there is a strictly positive \emph{fraction} of infected individuals in the population at the beginning of the epidemic. This corresponds to an infinite number of infecteds, no matter how small this fraction is. One way to obtain the setting of an epidemic with only few infectious individuals at the beginning is by starting with a small positive fraction of infected individuals at time $t=0$ (chosen in the right way), and then to let this fraction tend to zero while also letting time $t\to-\infty$. This is how we formulated~\eqref{eq:s}-\eqref{eq:FOI} without worrying too much about initial conditions. In the `far past' $t=-\infty$, the population is completely susceptible. If $R_0>1$ and we let $t\to+\infty$, then an epidemic outbreak will occur with final size characterized by~\eqref{eq:finalsize} (in case of a major outbreak). It is not the purpose of this paper to make these arguments fully rigorous. See also~\cite[Appendix A: The early stages']{Leung2016}. 

Together with the `far past' conditions, the deterministic model allow for the derivation of $R_0$ and the final size. By doing so, we find~\eqref{eq:R0} for $R_0$ and~\eqref{eq:finalsize} for the final size. In particular, the interpretation of $R_0$ and the final size is the same as in the stochastic reasoning. In fact, the law of large numbers results from~\cite{Ball1986} show that there is convergence of these epidemiological quantities of the stochastic process to the deterministic process as population size $n\to\infty$.

\section{Who is the infector?}\label{sec:cause}
\subsection{Timing matters}
Now that we have set up the model framework in Section~\ref{sec:model}, we can ask: who is the infector? We assume that a major outbreak occurs and the final number of infected individuals $Z\to\infty$. In particular, $R_0>1$. In the remainder of the text we take the deterministic viewpoint, while making use of probabilistic arguments. Assume that the final fraction of infected individuals $\bar Z$ converges to $z$, where $z$ is the nontrivial solution of~\eqref{eq:finalsize}. Recall from Section~\ref{sec:epi} that we let $\rho_a=1-\rho_d$ and $\rho_d$ denotes the fraction of $z$ that had a symptomatic infector. 

At time $t$, the rate at which new infections are caused by symptomatics is $F_d(t)s(t)$, and the rate at which new infections are caused by asymptomatics is $F_a(t)s(t)$. The fraction of individuals infected by symptomatic individuals is obtained by integrating over the entire epidemic. In particular, 
\begin{equation}\label{eq:rho_S}
\rho_d=\frac{\int_{-\infty}^\infty F_d(t)s(t)dt}{\int_{-\infty}^\infty (F_a(t)+F_d(t))s(t)dt} = \frac{\int_{-\infty}^\infty F_d(t)s(t)dt}{z}.
\end{equation}
Note that indeed $z=\int_{-\infty}^\infty (F_a(t)+F_d(t))s(t)dt$ (integrate both left- and right-hand side of~\eqref{eq:s} over all time and observe that $s(\infty)=1-z$, $s(-\infty)=1$). In principle, this characterizes $\rho_d$. A different characterization for $\rho_d$ involving the Malthusian parameter and (generally implicit) limiting random variables of branching processes can be found in~\cite[Theorem 2.7]{guilty}. However, in general, both the expression~\eqref{eq:rho_S} for $\rho_d$ and the alternative expression of~\cite{guilty} are not very informative and we have not managed to make either of them more explicit in this general setting. 

Due to the issues with obtaining general results for $\rho_d$ mentioned above, we instead derive general \emph{bounds} for $\rho_d$ in Section~\ref{sec:Bounds}. First, we obtain explicit expressions of $\rho_d$ for the very special case that $\pi_a(\tau)=\pi_d(\tau)$ for all $\tau$, i.e.\ asymptomatic and symptomatic cases have the same probability to be infectious at any given age-since-infection. From~\eqref{eq:FOI}, we find that 
\begin{equation*}
F_a(t)=\frac{(1-p_d)\lambda_a}{p_d\lambda_d}F_d(t).
\end{equation*}
Therefore, 
\begin{equation*}
z=\int_{-\infty}^\infty (F_a(t)+F_d(t))s(t)dt=\left(1+\frac{(1-p_d)\lambda_a}{p_d\lambda_d}\right)\int_{-\infty}^\infty F_d(t)s(t)dt,
\end{equation*} 
leading to
\begin{equation}\label{eq:rho_S_special}
\rho_d = \frac{p_d\lambda_d}{p_d\lambda_d+(1-p_d)\lambda_a} =\frac{p_dR_{0,d}}{R_0},
\end{equation}
where the second equality follows from $\Exp(\iota_a)=\Exp(\iota_d)$. Note that~\eqref{eq:rho_S_special} is monotonically increasing in both $p_d$ and $\lambda_d$. Furthermore, the fraction $\rho_d$ only depends on the type-specific reproduction numbers $R_{0,a}$ and $R_{0,d}$ and the probability $p_d$. In fact,~\eqref{eq:rho_S_special} is equal to the quantity $\theta_d$ (see~\eqref{eq:theta}) that indicates the importance of symptomatics at the beginning of an epidemic. 

Additionally, if also $\lambda_d=\lambda=\lambda_a$, the expression~\eqref{eq:rho_S_special} reduces even further to 
\begin{equation}\label{eq:rho_p}
\rho_d=p_d.
\end{equation}

\subsection{General upper- and lower bounds on $\rho_d$}\label{sec:Bounds}
Two features of $\rho_d$ seem obvious, but we have not been able to prove them since $\rho_d$ depends on the \emph{distributions} of the latent and infectious periods in a rather complicated way (see~\eqref{eq:rho_S}). Instead we conjecture the two features regarding $\rho_d$ and explain why we believe them to be true. First, we believe that $\rho_d$ is increasing in $p_d$ (and hence decreasing in $p_a=1-p_d$). This means that, keeping all other parameters fixed, the fraction infected by symptomatic individuals increases with the probability that an infected individual becomes a symptomatic case. Secondly, we believe that $\rho_d$ is increasing in $\lambda_d$ (and decreasing in $\lambda_a$). Note that in the special case that latent and infectious periods of symptomatics and asymptomatics are identical, i.e.\ $\pi_a(\tau)=\pi_d(\tau)$ for all $\tau$, we find from~\eqref{eq:rho_S_special} that these two statements regarding monotonicity of $\rho_d$ hold true. 

Note that as $p_d$ or $\lambda_d$ changes, the overall fraction infected $z$ also changes. We believe that when one keeps $R_0$ fixed (and therefore also $z$) as well as $R_{0,d}$, the statements regarding monotonicity of $\rho_d$ in $p_d$ and $\lambda_d$ remain true. If we change $\lambda_d$ while keeping $R_0$ and $R_{0,d}=\lambda_d\Exp(\iota_d)$ fixed, we also need to change the mean infectious period $\Exp(\iota_d)$. In the case that we change $p_d$, also $R_{0,a}$ changes. 

Even though we believe the monotonicity of $\rho_d$ in $p_d$ and $\lambda_d$ to hold, there is little qualitative insight that we can obtain from~\eqref{eq:rho_S}. Therefore, we derive explicit ranges within which $\rho_d$ will certainly lie, when keeping the reproduction numbers $R_0$, $R_{0,d}$, and $R_{0,a}$ fixed (and consequently also $p_d=1-p_a$), in particular since in many situations the exact distributions of latent and infectious periods are not known in detail. 

Now we provide an intuitive argument for our beliefs on the monotonicity of $\rho_d$ and the extreme model that provides us an upper bound $\rho_d^+$ (and for symmetry reasons also a lower bound $\rho_d^-$). Consider one newly infected symptomatic individual. Suppose we shorten the latent period of this individual to zero and we make its infectious pressure very intense (by  decreasing the infectious period and at the same time increasing its infectious contact rate so that the product remains constant), then this individual can cause infections earlier on in the epidemic when there are more susceptible individuals. So more contacts will result in infections, i.e.\ this symptomatic individual will cause more infections. By assuming that the new latent and infectious periods are deterministic, we ensure that the symptomatic individual actually cause more infections rather than only making it more probable (which would be the case with e.g.\ the Markov assumption). Moreover, increasing the latent period of an asymptomatic individual will imply that it can only start infecting others \emph{later} on in the epidemic. Note that by assuming that asymptomatics also have very intense infectious pressure, this also implies that the symptomatic secondary cases they generate become infectious earlier on in the epidemic as well. It seems reasonable that by applying the changes in latent and infectious periods to all individuals in the population, i.e.\ if we consider an extreme case of the class of continuous-time Reed-Frost models where $L_a$ is long relative to $L_d$, $\iota_a$, and $\iota_d$, \emph{and} very short $\iota_a$ and $\iota_d$ combined with very large $\lambda_a$ and $\lambda_d$, the extreme model will lead to an upper bound in $\rho_d$ for all models with fixed $R_0$, $R_{0,d}$, and $R_{0,a}$. Of course, all of this is by no means a proof (for the proof as well as the precise assumptions we need for the latent and infectious periods we refer to~\cite{guilty}), but we hope that this convinces the reader why $\rho_d^+$ is the upper bound (and/or encourages the reader to read~\cite{guilty}). 

The proof in~\cite{guilty} uses an epidemic graph construction that couples susceptibility processes with backward branching processes. The general idea in the proof presented in~\cite{guilty} is that there is a shortest (directed) path from the individual that is infectious at the beginning of the epidemic to the individual under consideration. Only the type of the neighbour with an edge pointing towards the individual under consideration determines whether it got infected by an asymptomatic or symptomatic case. We can determine this by considering the backward branching process of the individual under consideration. This branching process has an offspring distribution that is Poisson distributed with mean $p_d R_{0,d}$. The offspring of an individual $u$ are those individuals that have an edge pointing towards $u$ in the epidemic graph, i.e.\ those individuals that could potentially infect $u$ if they become infected themselves. The calculations in~\cite{guilty} provides us with the following upper bound $\rho_d^+$ (see also~\cite[Remark 4.1]{guilty}):
\begin{equation}\label{eq:rho_S_paper1}
\rho_d^+=\frac{\eta_d}{z}+\left(1-\frac{\eta_d}{z}\right)\frac{(2-\eta_d-z)p_dR_{0,d}}{2}.
\end{equation}
Here, $0\leq \eta_d<1$ is the largest solution to
\begin{equation}\label{eq:final1}
1-\eta_d=e^{-p_dR_{0,d}\eta_d},
\end{equation}
which is strictly between 0 and 1 if and only if $p_dR_{0,d}>1$ and zero otherwise. Moreover, $1-\eta_d$ can be interpreted as the probability that a randomly chosen individual is susceptible at the end of epidemic when only symptomatic cases can reproduce (set $\lambda_a=0$). We discuss $\eta_d$ in more detail in Section~\ref{sec:epicurve}.

The lower bound $\rho_d^-$ corresponding to~\eqref{eq:rho_S_paper1} is obtained by interchanging the roles of symptomatic and asymptomatic cases, i.e. 
\begin{equation}\label{eq:rho_S_lower}
\rho_d^-=1-\rho_a^+
\end{equation} 
where
\begin{equation}\label{eq:rho_a_upper}
\rho_a^+=\frac{\eta_a}{z}+\left(1-\frac{\eta_a}{z}\right)\frac{(2-\eta_a-z)(1-p_d)R_{0,a}}{2}.
\end{equation}
and $0\leq\eta_a<1$ is the largest solution to 
\begin{equation}\label{eq:finala}
1-\eta_a=e^{-(1-p_d)R_{0,a}\eta_a}
\end{equation}
(which is strictly between 0 and 1 if and only if $(1-p_d)R_{0,a}>1$ and zero otherwise). Note that~\eqref{eq:rho_S_paper1} and~\eqref{eq:rho_S_lower} are not explicit since $z$, $\eta_a$, and $\eta_d$ are only implicitly characterized through equations~\eqref{eq:finalsize}, \eqref{eq:final1}, and~\eqref{eq:finala}, respectively. Since $\rho_d^+$ and $\rho_d^-$ are upper- and lower bounds for all distributions of the latent and infectious periods with fixed $R_0$, $R_{0,a}$, and $R_{0,d}$ (with some technical restrictions, see~\cite{guilty} for details), we find that $\rho_d^- \leq \theta_d \leq \rho_d^+$. Indeed, $\rho_d=\theta_d$ in the special case that $\pi_a(\tau)=\pi_d(\tau)$.

\subsection{The epidemic curve for extreme settings}\label{sec:epicurve}
In this section we discuss the epidemic curve for the extreme setting where the mean latent period $\Exp(L_a)$ is long relative to $\Exp(L_d)$, $\Exp(\iota_a)$, and $\Exp(\iota_d)$. To characterize the curve, we revisit~\eqref{eq:final1} for $\eta_d$. Suppose that $p_dR_{0,d}>1$, then symptomatic individuals themselves can sustain an epidemic (without asymptomatic cases reproducing) and $0<\eta_d<1$ is unique and strictly positive. In fact, if $\Exp(L_a)$ is relatively long, asymptomatic cases will initially not reproduce, and we are in the setting of an epidemic outbreak generated by symptomatic individuals only. (Or, from a stochastic viewpoint, there is a positive probability that this occurs, and we condition on this happening.) 

What happens after the epidemic `symptomatic' outbreak? This very much depends on the distributions for $L_a$, $L_d$, $\iota_a$, and $\iota_d$! In fact, we discuss the epidemic curves for the class of continuous time Reed-Frost models and Markov models in Sections~\ref{sec:constant} and~\ref{sec:Markov} below. The upshot is that these two classes of models exhibit very different behaviour. Note that, when $p_d R_{0,d}<1$, there is no initial `symptomatic' outbreak, and the epidemic curve takes the form of a `standard' curve. 

\subsubsection{The continuous-time Reed-Frost model}\label{sec:constant}
In the continuous-time Reed-Frost model, latent and infectious periods are deterministic. Therefore, we know with certainty that newly infected asymptomatic individuals will  initially remain in their latent period and do not transmit. There is a `symptomatic' outbreak until there are no more infectious symptomatics left. At this point, when the `first symptomatic wave' of the epidemic has occurred, the probability that a randomly chosen individual is susceptible is given by $1-\eta_d$. Note that this allows us to interpret the fraction $\eta_d/z$ in~\eqref{eq:rho_S_paper1} as the probability that an individual was infected in the first wave of the epidemic, given that it got infected. 

After the first wave, the epidemic continues when the asymptomatic cases of the first outbreak become infectious. Additionally, there will be newly infected symptomatic cases. Together these infectives generate a second wave of infections (where the asymptomatic cases generated in the second wave are not yet infectious). The same reasoning as before yields the probability that an individual escapes infection in the second wave, given that it escaped infection in the first wave, is $(1-\eta_2)/(1-\eta_d)$. Here $\eta_2$ denotes the fraction of infectives after the second wave (so $\eta_1:=\eta_d$ would be consistent notation). Such an individual needs to escape infection from the $(1-p_d)\eta_d$ asymptomatic individuals of the first wave as well as all the symptomatically infected individuals from the second wave. Hence $0<\eta_2<1$ is the unique solution to
\begin{equation}\label{eq:final2}
\frac{1-\eta_2}{1-\eta_d}=e^{-\big((1-p_d) R_{0,a} \eta_d+p_dR_{0,d}(\eta_2-\eta_d)\big)}.
\end{equation}
Note that, contrary to the first wave, the second wave is not an epidemic \emph{outbreak} per se. Indeed, it starts with a significant fraction $(1-p_d)\eta_d$ of latently infected individuals. The susceptible population is of size $1-\eta_d$, and even if at the end of the second wave we have a fraction $\eta_2$ infected individuals, we may have $R_0(1-\eta_d)<1$.

By the same reasoning, we find that 
\begin{equation}\label{eq:finalk}
\frac{1-\eta_k}{1-\eta_{k-1}}=e^{-\big((1-p_d) R_{0,a} (\eta_{k-1}-\eta_{k-2})+p_dR_{0,d}(\eta_k-\eta_{k-1})\big)},
\end{equation}
for $k\geq3$, where $0<\eta_k<1$ is the fraction of individuals infected in one of the first $k$ waves (and $\eta_1=\eta_d$). So, we find that the fraction of susceptibles depletes over time to $1-z$ in an infinite number of waves that are characterized by $1-\eta_d$, $1-\eta_2$, $1-\eta_3,\ldots$ This qualitative multi-wave process is illustrated in {Figs.~\ref{fig:multiwave} and~\ref{fig:2wave}} of Section~\ref{sec:numerics}.

\subsubsection{The Markov model}\label{sec:Markov}
The crucial difference between the continuous-time Reed-Frost models and the Markov models is that we now only know that \emph{on average} newly infected asymptomatic individuals will remain in their latent period during the `symptomatic outbreak' since $\Exp(L_a)$ is relatively long. 
However, due to the memoryless property of the exponential distribution, there is a constant probability per unit of time that an asymptomatic case in the latent stage becomes infectious, regardless of how long ago that individual got infected (in contrast to the continuous-time Reed-Frost of Section~\ref{sec:constant} with deterministic latent periods). 

Given that a major outbreak occurs, at first, the probability that any of the asymptomatic cases become infectious in the `symptomatic outbreak' is small enough to neglect. Therefore, as in Section~\ref{sec:constant}, the epidemic starts with a symptomatic outbreak that is characterized by $\eta_d$ satisfying~\eqref{eq:final1}. But, after the first `symptomatic' wave, there is a significant fraction $(1-p_d)\eta_d$ of asymptomatic individuals in the latent stage. These all have a constant probability per unit of time to become infectious (whereas in Section~\ref{sec:constant} these asymptomatics became infectious more or less at the same time). As a result, after the first (symptomatic) wave, both symptomatic and asymptomatic cases can transmit infection to the remaining susceptible population. The susceptible fraction slowly decreases from $1-\eta_d$ down to the fraction $1-z$. The time scale at which this occurs is of the same order as $\Exp(L_a)$. In particular, the epidemic curve can be approximated by a two-wave process rather than the multi-wave process of Section~\ref{sec:constant}. This qualitative two-wave process is {illustrated} in Fig.~\ref{fig:2wave} of Section~\ref{sec:numerics}.

\section{Numerical illustration}\label{sec:numerics}
We consider different scenarios for the parameter values to illustrate the analytical results of this paper. We use the system of ODE~\eqref{eq:ODE_exp} and the system of DDE~\eqref{eq:DDE_RF} for the numerical simulations in this section. 

\subsection{Epidemic curves: Markov models vs.\ continuous-time Reed-Frost models}\label{sec:extreme_numerics}
We choose parameter values to illustrate (i) the extreme behaviour described in Sections~\ref{sec:constant} and \ref{sec:Markov} and (ii) the difference between the Markov (M) assumption and the Reed-Frost (RF) assumption in temporal behaviour. We let $R_0=2$, $R_{0,d}=2.5$, $p_d=0.5$ (consequently $R_{0,a}=1.5$). Note that the final sizes $z$ and $\eta_d$, the lower- and upper bounds $\rho_d^-$ and $\rho_d^+$, and the fraction $\theta_d$ of $R_0$ caused by symptomatic cases depend only on $R_0$, $R_{0,d}$, and $p_d$. The values of these epidemiological quantities for scenarios 1 and 2 are summarized in Table~\ref{table:epi}. 

\begin{table}[H]
\centering
\begin{tabular}{l|l|l|l|l|l}
Epidemiological quantity & $z$ & $\eta_d$ & $\rho_d^-$ & $\rho_d^+$ & $\theta_d$ \\
\hline 
Value & 0.80 & 0.37 & 0.55 & 0.74 & 0.63
\end{tabular}  
\caption{Epidemiological quantities evaluated for the parameter values $R_0=2$, $R_{0,d}=2.5$, and $p_d=0.5$ ($R_{0,a}=1.5$).}
\label{table:epi}
\end{table}

\begin{figure}[H]
\centering
\includegraphics[scale=0.6]{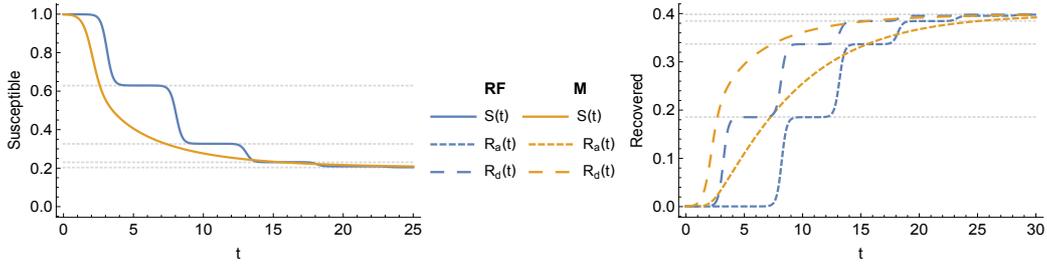}
\caption{Scenario 1 for the class of Markov models (M, yellow) and the class of continuous-time Reed-Frost models (RF, blue). Parameter values are $R_0=2$, $R_{0,d}=2.5$, $p_d=0.5$ (consequently $R_{0,a}=1.5$), $\Exp(L_a)=5$, $\Exp(L_d)=0$, $\Exp(\iota_a)=0.10=\Exp(\iota_d)$. }
\label{fig:multiwave}
\end{figure}

\begin{figure}[H]
\centering
\includegraphics[scale=0.6]{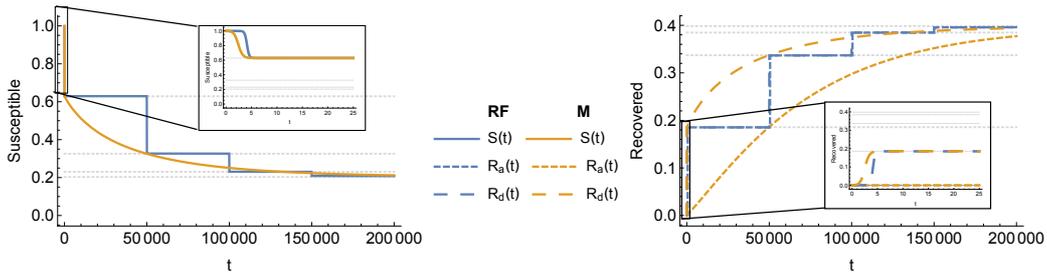}
\caption{Scenario 2 for the class of Markov models and the class of continuous-time Reed-Frost models. The inset figures zoom in on $t\in(0,25)$. Parameter values are $R_0=2$, $R_{0,d}=2.5$, $p_d=0.5$ (consequently $R_{0,a}=1.5$), $\Exp(L_a)=50000$, $\Exp(L_d)=0$, $\Exp(\iota_a)=0.10=\Exp(\iota_d)$.}
\label{fig:2wave}
\end{figure}

When it comes to the time evolution of the epidemic, the mean values and the distribution of $L_a$, $L_d$, $\iota_a$, and $\iota_d$ start to matter as we will show using the M and RF assumption. We consider the following two scenarios. We let $\Exp(L_d)=0$, $\Exp(\iota_a)=0.10=\Exp(\iota_d)$. Scenarios 1 and 2 differ only in the mean latent period $\Exp(L_a)$ for asymptomatic cases: $\Exp(L_a)=5$ for scenario 1 and $\Exp(L_a)=50000$ for scenario 2. The reason for choosing such big differences in mean latent periods between symptomatics and asymptomatics is to clearly illustrate the behaviour described in Section~\ref{sec:epicurve}. Furthermore, the scenarios are such that the RF model is close to the extreme model that yields the upper bound $\rho_d^+$ for all classes of models with the same $p_d$, $R_0$, and $R_{0,d}$. The epidemic curves for the two scenarios under the M and RF assumption are found in Figs.~\ref{fig:multiwave} and~\ref{fig:2wave}. 

Note that, in both classes of models M and RF, according to theory, the fraction of susceptibles monotonically decreases over time and reaches the level $1-z$, with $z$ the final fraction of infecteds in the epidemic. Furthermore, in Fig.~\ref{fig:multiwave}, we find that the RF model displays the extreme behaviour as described in Section~\ref{sec:constant}. The level $1-\eta_k$ of the $k$th wave is indeed characterized by the equations~\eqref{eq:final1}-\eqref{eq:finalk}. Next, we also note that the M model does \emph{not} display any extreme behaviour under scenario 1. In other words, the mean latent period $\Exp(L_a)$ is not long enough (relatively speaking). In particular, we find that the two classes of models display very different behaviour, i.e.\ a `multi-wave epidemic' and a `standard epidemic'. In Fig.~\ref{fig:2wave} we consider scenario 2 and both the RF and M model exhibit the extreme behaviour described in Sections~\ref{sec:constant} and~\ref{sec:Markov}.

Evaluating $\rho_d$ numerically yields 0.73 (M) and 0.74 (RF) for scenario 1 and 0.74 (M and RF) for scenario 2. The upper bound (for both scenarios and classes of models) is $\rho_d^+=0.74$; see Table~\ref{table:epi}. Hence we find that the two scenarios are both extreme enough that they yield $\rho_d$ close to $\rho_d^+$. We find that there is a significant difference between $\rho_d$ and $\theta_d=0.63$, i.e.\ symptomatic cases play a different role for the beginning of an outbreak as in the final size of the epidemic. Furthermore, when it comes to $\rho_d$, the M or RF assumption has little effect in the two scenarios that we consider (this need not be true in general, see Section~\ref{sec:rhod_numerics}). But, we also find that the time evolution of the epidemic are very different between the Markov and Reed-Frost models (Figs.~\ref{fig:multiwave} and~\ref{fig:2wave}).

\subsection{Numerical investigation of $\rho_d$}\label{sec:rhod_numerics}
\begin{figure}[H]
\centering
\includegraphics[scale=0.55]{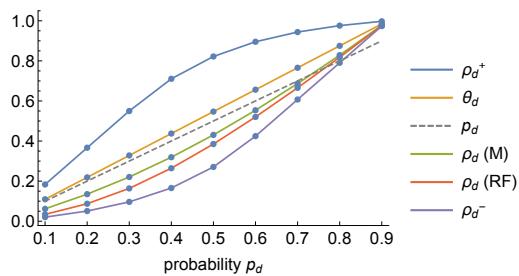}
\caption{Varying $p_d$ for $R_0=3.2$, $R_{0,d}=3.5$ (and $R_{0,a}$ is changed using~\eqref{eq:R0}). Mean latent and infectious periods are $\Exp(L_a)=1$, $\Exp(L_d)=5$, $\Exp(\iota_a)=1.4$, $\Exp(\iota_d)=2$.}
\label{fig:rhod_example}
\end{figure}

We consider some numerical examples where parameter values are chosen to illustrate our beliefs on the monotonicity of $\rho_d$ that we stated in Section~\ref{sec:Bounds}. We fix $R_0=3.2$, $R_{0,d}=3.5$, and we vary $p_d$. These parameter values yield lower- and upper bounds $\rho_d^-$ and $\rho_d^+$ as functions of $p_d$ as well as parameter $\theta_d$ in between the bounds; see Fig.~\ref{fig:rhod_example}. Fig.~\ref{fig:rhod_example} illustrates that the range between $\rho_d^-$ and $\rho_d^+$ can be quite wide. Next, we consider $\rho_d$ for the M and RF models. We let $\Exp(L_a)=1$, $\Exp(L_d)=5$, $\Exp(\iota_a)=1.4$, and $\Exp(\iota_d)=2$. The example of Fig.~\ref{fig:rhod_example} supports our first general belief, that $\rho_d$ is an increasing function of $p_d$. 

\begin{figure}[H]
\centering
\includegraphics[scale=0.55]{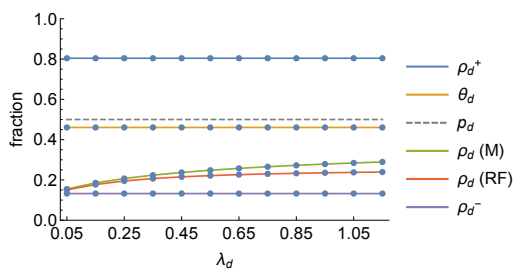}
\caption{Varying $\lambda_d$ for $p_d=0.5$, $R_0=3.8$, $R_{0,d}=3.5$ (and $R_{0,a}=4.1$). Mean latent and infectious periods are $\Exp(L_a)=1$, $\Exp(L_d)=5$, $\Exp(\iota_a)=1.4$ (and $\Exp(\iota_d)$ varies according to~\eqref{eq:R0}).}
\label{fig:rhod_example2}
\end{figure}

For our second example, we fix $p_d=0.5$, $R_0=3.8$, $R_{0,d}=3.5$, $\Exp(L_a)=1$, $\Exp(L_d)=5$, $\Exp(\iota_a)=1.4$, and vary $\lambda_d$ (consequently we also vary $\Exp(\iota_d)$); see Fig.~\ref{fig:rhod_example2} for the result. Note that also $R_{0,a}$ is fixed (at 4.1). Then $\rho_d^+$, $\rho_d^-$, $\theta_d$, and $p_d$ are constants as a function of $\lambda_d$. On the other hand, in this example, $\rho_d(M)$ and $\rho_d(RF)$ are both monotonically increasing functions of $\lambda_d$, which supports our second conjecture. 

\subsection{Illustration with norovirus and measles}
We take two scenarios `norovirus' and `measles' that represent real-life epidemics with parameter values in reasonable ranges taken from literature. This section serves to illustrate the role of $\rho_d$. By no means do we try to incorporate the complexity of the two infections. 

\begin{table}[H]
\centering
\begin{tabular}{l|l|l|l|l|l|l|l|l}
Scenario & $R_0$ & $R_{0,d}$ & $R_{0,a}$ & $p_d$ & $\Exp(L_a)$ & $\Exp(L_d)$ & $\Exp(\iota_a)$ & $\Exp(\iota_d)$ \\
\hline
Norovirus & 1.40 & 1.64 & 0.85 & 0.70 & 1.5 & 1.5 & 19.3 & 25.3 \\
Measles & 2.47 & 17 & 0.85 & 0.1 & 7 & 7 & 0.5 & 5.0
\end{tabular}  
\caption{Parameter values for the two scenarios `norovirus' and `measles'.}
\label{table:parameters_inf}
\end{table}

The first scenario represents a norovirus outbreak in a hospital setting, and we take reference values based on~\cite{Sukhrie2012, Newman2016}. From~\cite{Newman2016}, we obtain $p_d=0.7$, duration of shedding $\Exp(\iota_a)=19.3$ days and $\Exp(\iota_d)=25.3$ days, and mean latent periods $\Exp(L_a)=1.5 \text{ days}=\Exp(L_d)$. \cite{Sukhrie2012} report estimated reproduction numbers of $R_{0,a}=0.85$ and $R_{0,d}=1.64$ for asymptomatic and symptomatic cases respectively. Furthermore, from their data, we find $p_d=0.7$ consistent with~\cite{Newman2016}. Hence $R_0=1.38$.  

The second scenario represents measles in England in the vaccine era with subclinical infections. We base parameter values for the reproduction numbers, $p_d$, and mean infectious periods on~\cite[Fig.~1]{Glass2004}. Unvaccinated individuals that are infected are symptomatic and have a high reproduction number. Vaccinated individuals could potentially become asymptomatically infected but these have a very short infectious period and low reproduction number (in~\cite{Glass2004} the possibility of asymptomatic infection is related to antibody decay). The model in~\cite{Glass2004} is used to address a completely different question related to waning and boosting of immunity, and does not include any latent period. For our purposes we let mean latent periods be $\Exp(L_a)=7 \text{ days}=\Exp(L_d)$, coinciding with a reasonable incubation period~\cite{CDC2018}. Again, we do not try to fit our model to any real measles outbreak, but rather want to illustrate the use of the quantity $\rho_d$ by considering reasonable parameter values.

\begin{figure}[H]
\centering
\includegraphics[scale=0.45]{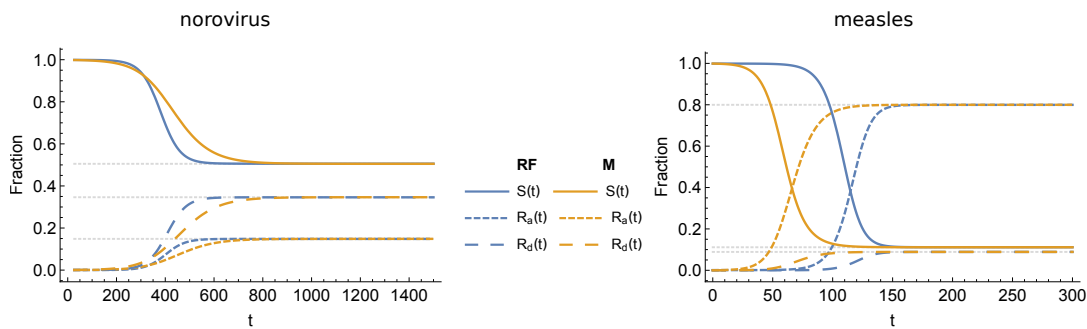}
\caption{Illustration of the evolution of the fraction of susceptible and recovered individuals for parameter values representing the scenarios norovirus, and measles for the class of M and RF models. Note the time scales of the epidemic outbreaks for our simple models are not representative for real outbreaks. Parameter values are summarized in Table~\ref{table:parameters_inf}.}
\label{fig:numerics}
\end{figure}

Parameter values for both scenarios are summarized in Table~\ref{table:parameters_inf}. Note that both scenarios have $R_0>1$ and $p_d R_{0,d}>1$. The resulting epidemic curves are presented in Fig.~\ref{fig:numerics} for both the M and the RF model. Note that, for both scenarios, the curves of both the M and RF model are of a `standard epidemic'. Qualitatively, the two curves are similar. However, when it comes to the quantitative dynamics, we do find that the assumption on the distribution of the latent and infectious periods matter.

Next, we consider the fraction $\rho_d$ of cases that were caused by symptomatic individuals, given that they were ultimately infected in Table~\ref{table:rho_S}. Note that lower- and upper bounds $\rho_d^-$ and $\rho_d^+$ are only dependent on $p_d$, $R_0$, and $R_{0,d}$ and are calculated using~\eqref{eq:rho_S_lower} and~\eqref{eq:rho_S_paper1}. We find that (i) for both the scenarios that the RF and the M model assumption yield similar values for $\rho_d$ (ii) the upper- and lower-bounds $\rho_d^+$ and $\rho_d^-$ are narrow for the norovirus scenario but wider for the measles scenario, and (iii) for both scenarios $\rho_d$ and $\theta_d$ are close to each other for both the RF and the M model assumption. Furthermore, while the measles scenario has only a small proportion $p_d=0.1$ of infected individuals that become symptomatic, both in the beginning of the epidemic and in the final size, they play a much bigger role than the asymptomatic cases. Indeed, we find that symptomatic cases contribute between $\rho_d^-=0.57$ and $\rho_d^+=0.86$ of all transmissions that occur in the epidemic. In the norovirus scenario, this difference in the fractions $p_d$ and $\rho_d$ is not as extreme. Still we find that the contribution of symptomatic cases to the transmissions is between $\rho_d^-=0.81$ and $\rho_d^+=0.85$, which is larger than would be expected based on $p_d=0.7$ alone. We find that symptomatic cases cause most of the infections in the norovirus scenario. This is consistent with the findings of~\cite{Sukhrie2012} that symptomatic cases contributed most to the spread of infection in the outbreak settings that they analysed (incidentally, the outbreak duration is not consistent with~\cite{Sukhrie2012}).

\begin{table}[H]
\centering
\begin{tabular}{l|l|l|l|l|l|l|l}
Scenario & $z$ & $\eta_d$ & $\rho_d^-$ & $\rho_d$ (RF) & $\rho_d$ (M) & $\theta_d$ & $\rho_d^+$ \\
\hline
Norovirus & 0.49 & 0.25 & 0.81 & 0.82 & 0.82 & 0.82 & 0.85 \\
Measles & 0.89 & 0.69 & 0.57 & 0.67 & 0.65 & 0.69 & 0.86 
\end{tabular}  
\caption{The fraction $\rho_d$ of the final size $z$ that is caused by symptomatic cases for norovirus and measles. $\rho_d^-$ and $\rho_d^+$ denote the lower- and upper bound for $\rho_d$, $\rho_d$ (RF) and $\rho_d$ (M) are the numerical approximations of $\rho_d$ under the RF and M assumption, and $\theta_d$ is the fraction of secondary cases at the beginning of the outbreak that are caused by symptomatic cases.}
\label{table:rho_S}
\end{table}

\section{Conclusion and discussion}\label{sec:conclusion}
In this paper we considered an SEIR model with symptomatic and asymptomatic cases and looked at the fraction $\rho_d$ of the infected individuals that was caused by symptomatic cases. The quantity $\rho_d$ gives an indication of the importance of symptomatic cases in the transmission process in an epidemic. In general, it is hard to make statements about $\rho_d$ as timing of transmission events matter: a susceptible individual that gets infected by an asymptomatic case can no longer become infected by a symptomatic case. We were only able to derive an explicit expression for $\rho_d$ for the simple situation where the latent and infectious period distributions of symptomatic and asymptomatic coincide. 

Instead we used this timing issue to our advantage. We considered extreme settings for the latent and infectious periods of symptomatics and asymptomatics. The current paper is accompanied by a twin paper~\cite{guilty} in which the arguments and reasoning are made precise (in a more general setting than we consider here). The extreme setting allowed us to derive an upper and lower bound $\rho_d^+$ and $\rho_d^-$ for $\rho_d$. The bounds $\rho_d^+$ and $\rho_d^-$ are the bounds for fixed $R_{0,d}, R_{0,a}$ and $p_d$ but unspecified latent and infectious periods. Although the expressions for $\rho_d^+$ and $\rho_d^-$ are still implicit (depending on some implicit final size equations) they are relatively simple and easily obtained. We studied $\rho_d$ and the bounds numerically in Section~\ref{sec:numerics}. It was seen that the biggest range of possible values of $\rho_d$, i.e.\ the biggest difference between $\rho_d^-$ and $\rho_d^+$, is for the situation where both symptomatics and asymptomatics separately have big enough reproduction numbers to produce an outbreak on their own: $p_dR_{0,d}>1$ and $p_aR_{0,a}>1$ as seen in Fig.~\ref{fig:rhod_example}.

We stated two rather natural beliefs for the general model, namely that $\rho_d$ increases with increasing fraction $p_d$ of infecteds that become symptomatic, and that $\rho_d$ increases with increasing infectious contact rate $\lambda_d$ from symptomatic cases. Unfortunately we were not able to prove either statement, also not using the techniques employed in~\cite{guilty}, at least not in a straightforward way. Instead we pose these two statements as open problems for future work.

We paid special attention to two special classes of models, namely the class of Markov models (with latent and infectious periods exponentially distributed) and continuous-time Reed-Frost models (with deterministic latent and infectious periods). We specifically considered the epidemic curve in the extreme setting with fixed mean latent and infectious periods in Sections~\ref{sec:constant} and~\ref{sec:Markov} and found large differences in the qualitative behaviour of the two classes of models. In other words, not only the means but also the distributions matter! At this point, we hope that Sections~\ref{sec:constant} and~\ref{sec:Markov} and Figs.~\ref{fig:multiwave} and~\ref{fig:2wave} have convinced readers that the choice of distribution functions matter. This fact is of course not new and has been observed before in different settings (e.g.~\cite{Diekmann2013,Omori2015,Pellis2015,Mees11,Vergu2010,Kuul82}). The distribution of the infectivity profile can matter a lot. However, how much it matters depends on the aspects of the epidemic model that one considers. In case of $R_0$, it is the \emph{mean} infectious period that matters and not its distribution. In case of the Malthusian parameter (or as we have seen in this paper, the epidemic curve), the choice of distributions can matter a lot. Assuming exponential distributions for mathematical convenience could prove dangerous when not also considering the possible influences of this assumption. The qualitative outcome can be dramatically different depending on other choices. The critical reader will now point out that parameter values of Figs.~\ref{fig:multiwave} and~\ref{fig:2wave} probably do not fit any reasonable known disease. This is a valid criticism. However, our first aim was to illustrate possible effects. Moreover, much more reasonable parameter values can be chosen such that the assumption of constant periods create a multi-wave epidemic while the Markov assumption yields a `standard' epidemic curve (Fig.~\ref{fig:multiwave}). Finally, while one can argue about the realism of the simple models and parameter values used in this text, we hope to have reminded the reader that conclusions are always based on the assumptions that one makes. Furthermore, we hope one keeps in mind that it is worth exploring what the possible differences are by assuming something other than exponential distributions. 

Studying the end of the epidemic and the question of who the infector was is more involved than studying $R_0$ for the beginning of the epidemic. We managed to gain insights by providing bounds. This increased understanding of the transmission dynamics may help us in deciding the role that asymptomatic cases can play, and how their role may differ throughout an epidemic outbreak.

\subsection*{Acknowledgements}
The authors are supported by the Swedish Research Council (VR) [grant 2015-05015 (TB and KYL) and grant 2016-04566 (PT)].


\end{document}